\documentstyle[12pt]{article}
\def\lsim{\raise0.3ex\hbox{$<$\kern-0.75em\raise-1.1ex\hbox{$\sim$}}}
\def\gsim{\raise0.3ex\hbox{$>$\kern-0.75em\raise-1.1ex\hbox{$\sim$}}}
\begin {document}
\begin{center}
{\Large {\bf AN ESTIMATE OF THE PERCOLATION PARAMETER IN HEAVY ION
COLLISIONS}}
\\
\vskip 1.5 truecm
{\bf J.Dias de Deus and Yu. M. Shabelski$^*$}\\
\vskip 0.5 truecm
CENTRA, Instituto Superior T\'ecnico, 1049-001 Lisboa, Portugal
\vskip 0.75 truecm
\end{center}
\vskip 1.5 truecm
\begin{center}
{\bf ABSTRACT}
\end{center}

From existing hadron and heavy ion collisions data on $\bar{p}/p$
central production we estimate the value of the percolation parameter at
RHIC energies.

\vskip 1.5 truecm

\noindent $^*$Permanent address: Petersburg Nuclear Physics Institute,
Gatchina, St.Petersburg, Russia \\
E-mail: shabelsk@thd.pnpi.spb.ru

\newpage

It is well-known that in high energy hadron-nucleus collision there exists
inelastic screening \cite{Gr,KM} experimentally confirmed, especially for
the case of hadron-deuteron interactions. The same inelastic screening
has to exist in high energy heavy ion collision, as well. This
effect is very small for integrated cross sections (because many of them
are determined by geometry), but it is very important \cite{CKT} for the
calculations of secondary multiplicities and inclusive densities.
Similar results are obtained \cite{JU,APS} in the framework of string
fusion \cite{ABP}, or percolation \cite{BP} models, where string
fusion/percolation effects directly correspond \cite{APSh} to pomeron
interactions and are responsible for the suppression of particle
production.

The effect of percolation for central heavy ion collision is, in good
approximation determined by the reduction factor \cite{BP1}
\begin{equation}
F(\eta) = \sqrt{\frac{1-\exp^{-\eta}}{\eta}} ,
\end{equation}
$\eta$ being the transverse density parameter,
\begin{equation}
\eta = \frac{r^2_s N_s}{R^2} ,
\end{equation}
where $r_s$ is the string transverse radius (phenomenological estimation
gives $r_s \sim 0.2-0.3 fm$ \cite{ABFP}), $R^2$ the square of nuclear
overlapping which for central collisions is equal to nuclear radius
squared and $N_s$ is the number of produced strings. At $\eta \to 0,
F(\eta) \to 1$ (no percolation/inelastic screening) and at $\eta \to
\infty, F(\eta) \to 1/\sqrt{\eta}$ (maximal screening). The detailed
discussion of these behaviours can be found in \cite{BP1}.

In the present paper we will give an estimate for the percolation
parameter $\eta$ from  experimental data, not all of them connected to
heavy ion physics.

Let us, in fact, start from $\gamma p$ collisions at $W \sim 200$ GeV
(HERA). In lab. frame the asymmetry between comparatively slow $p$ and
$\bar{p}$ was observed to be \cite{H1}
\begin{equation}
A_B = 2 \frac{N_p-N_{\bar{p}}}{N_p+N_{\bar{p}}} =
(8.0 \pm 1.0 \pm 2.5) \% .
\end{equation}
That corresponds to the yield ratio
\begin{equation}
R^{\gamma p} = N_{\bar{p}}/N_p = 0.92 \pm 0.03 .
\end{equation}
However, the HERA kinematics is an asymmetrical one and particles rather
slow in lab. HERA frame are rather fast in c.m. frame. To account for this
we can use the Quark-Gluon String Model (QGSM) \cite{KTM,KTMSh} with
string junction diffusion, see details in \cite{ACKSh}. This correction is
not numerically large, the model estimation for c.m. photon-proton frame
gives $R^{\gamma p} = 0.86 \pm 0.02$.

In $\gamma p$ collisions we have the baryon number flux from one proton.
In the case of $pp$ interactions this flux should be two times larger,
that corresponding to
\begin{equation}
R^{pp} = 0.72-0.76
\end{equation}
in c.m. $pp$ frame.

The ratio $R^{AuAu}$ in central region of $AuAu$ collisions was
measured at energy 200GeV per nucleon at RHIC. The values are $0.74 \pm
0.02 \pm 0.03$ \cite{PHOBOS} and $0.75 \pm 0.04$ \cite{BRAMS}, i.e.
practically the same as we obtain for $pp$ collisions. This is in
principle unexpected as the sea contribution, which is $\bar{p}p$
symmetrical, is much more important in heavy ion collisions. The string
fusion/screening argument gives a fair explanation of what happens.

Now let us note that in the QGSM as well as in the Dual String Model (DSM)
\cite{JU,JU:2} the multiplicity of every secondary $h$ produced in $pp$
collision can be written as
\begin{equation}
<n^{pp}_h> = V_h + S_h ,
\end{equation}
where $V_h$ and $S_h$ are the contributions of valence and sea quarks,
respectively.

In the case of heavy ion collisions the correspondent equation can be
written as \cite{AP}

\begin{equation}
<n^{AA}_h> = N_{pair}V_h + (<\nu>_{AA}-N_{pair})F(\eta)S_h ,
\end{equation}
where $N_{pair}$ is the number of pair of nucleon-nucleon interactions and
$<\nu>_{AA}$ is the total number of binary interactions. $F(\eta)$
accounts for the sea strings percolation effect, a negligible effect for
$pp$ collisions.

The structure of this Eq. is rather evident, for equal nuclei we have
$N_{pair}$ interactions of valence quarks and diquarks (valence-valence
strings) and all another $<\nu>_{AA} - N_{pair}$ are sea quark
interactions (sea-sea strings) by definitions. Of course it is true only
in average, in every separate event the situation can be more
complicated because the number of interacting nucleons in every nucleus
can be different.

We account the percolation factor in Eq.(7) only for sea string
contribution due to three reasons. First of all, the number of sea
strings is in our case about 5 times larger than the number of valence
strings, so percolation effects for sea strings are more important.
Second and more physically important reason is that valence-valence
strings (contribution $V_h$ in Eq. (7)) are long in rapidity space
whereas sea-sea strings are rather short and their ends are distributed
more close to the central region. The suppression for the inelastic
screening effects coming from nuclei form factors (see [3] for details)
allows the total fusion of sea-sea strings but valence-valence strings
can be fused only in part, so the fusion/percolation effects will be
important for the last ones only at very high energies. Third, more
model dependent reason is that in DTU approach the interacting of
valence quarks is considered as a first step, and after several sea
quark interactions can be added, if necessary. So, the screening
effects should decrease firstly the number of sea-sea strings.

In order to have $R^{pp}/R^{AA} \approx 1$ at RHIC energies, as
experimentally observed, the ratio between valence quark and sea quark
contributions in Eqs. (6) and (7) should be approximately the same. So we
have
\begin{equation}
\frac{(<\nu>_{AA} - N_{pair}) F(\eta)}{N_{pair}} \simeq 1 .
\end{equation}
Using the experimental estimates of $<\nu>_{AA}$ and $N_{pair}$ from
\cite{PHENIX} for high-$p_T$ hadron production (these values are in
agreement with standard Glauber-like estimates \cite{JSh}) we obtain for
the most central collisions the value $F(\eta) = 1/4.8$, giving
\begin{equation}
\eta = 23 .
\end{equation}
Due to experimental errors and some model calculations we estimate
the error bar in the last value to be a factor of the order 1.5.
This result agrees with an estimate of $\eta$, making use of $pp$ central
charged particle densities at $\sqrt{s} \simeq$ 200 GeV, $\eta \simeq 16$
(see \cite{JU:2}). The value obtained for $\eta$ shows that we are well
inside the percolation (perhaps quark-gluon plasma) region. Note that if
$R^{pp} > R^{AA}$, as it would have happened without the centre of mass
correction bringing $R^{\gamma p}$ down, from = 0.92 to 0.86, then the
$\eta$ value should have been even larger.

The contributions of valence and sea strings in QGSM for $pp$ collisions
are of the same order. For $AuAu$ central interaction the sea contribution
is enhanced by the factor
\begin{equation}
\frac{<\nu>_{AA} - N_{pair}}{N_{pair}} ,
\end{equation}
which is, without percolations, about 4.8. That should increase the
inclusive density of secondaries. The suppression factor $F(\eta)$ about
1/5 for sea contribution decreases the total rapidity density in central
region about 2.5 or 3 times, that is in agreement with the previous
calculations of refs. \cite{CKT,APS,JShU}.

In our approach, the reason why the ratio $\bar{p}p$ is not much larger
in heavy ion collisions is caused by the same mechanism that limits
particle density in central rapidity: string fusion. It would be
interesting to see if saturation models \cite{Sat,Sat1} can also obtain
a similar effect in $\bar{p}p$ suppression.

We are grateful to N.Armesto for discussions.

\vskip 0.5 truecm


\end{document}